# Discovery of niobium hydride precipitates in superconducting qubits

Jaeyel Lee[1†], Zuhawn Sung[1†], Akshay A. Murthy[1†], Matthew J. Reagor[2], Anna Grassellino[1], and Alexander Romanenko[1*]

[1]Superconducting Quantum Materials and Systems Center, Fermi National Accelerator Laboratory; Batavia, IL 60510, USA

[2]Rigetti Computing; 775 Heinz Avenue, Berkeley, CA 94710, USA

[†] These authors contributed equally

*Corresponding Author: aroman@fnal.gov

**Abstract**

Superconducting qubits are at the forefront of quantum computing architectures but suffer from a relatively low coherence as compared to some other platforms (i.e. ion traps). Here we reveal the presence of niobium hydrides within niobium parts of 2D superconducting qubits – a previously unknown source of decoherence. We use complementary techniques of transmission electron microscopy (TEM), atomic force microscopy (AFM), and time-of-flight secondary ion mass spectroscopy (TOF-SIMS) to directly observe the niobium hydride precipitates and hydrogen distribution in full qubit chips both at room and cryogenic temperatures. We also identify possible sources of hydrogen and hydrides and pathways for improving the coherence by eliminating them.

**One-Sentence Summary:** We reveal the widespread niobium hydride precipitates in nominally niobium film parts of superconducting qubits – a previously unknown source of decoherence and coherence variability between cooldowns in qubits.

Niobium thin films of ~100-200 nm thickness are key components of many superconducting qubit architectures for quantum computing applications, where they are used to form the readout resonators, coupling lines, and the capacitance pads present in qubits (*1, 2*). Any microwave dissipation mechanisms present in these niobium films are therefore primary contributors to the decoherence of the superconducting qubits. There have been intensive studies to understand the microscopic origins of the decoherence in the superconducting qubits (*1, 3*), and several coherence-limiting mechanisms have been identified. These include two-level systems (TLS) hosted by amorphous oxides at various interfaces (*1, 4, 5*), i.e. metal-air (*6-8*), metal-substrate (*9-11*), and substrate-air interfaces (*6, 10*), which have been recently shown to be the primary coherence-limiting mechanism in the quantum regime for 3D superconducting radio frequency (SRF) cavities (*12*). Additionally, non-equilibrium quasiparticles (*13*), including those produced by cosmic rays and radioactive sources in the environment (*14, 15*) are considered as another possible decoherence mechanism.

It has been previously discovered in bulk niobium SRF cavities (*16-19*) that the formation of the non-superconducting niobium nano-hydrides upon cooldown from room temperature to cryogenic temperatures (<~150K) introduces a major dissipation mechanism, which limits achievable quality factors (*18*). This process is enabled by the fact that Nb can absorb a significant amount of hydrogen (H) even at room temperature - whenever it is not passivated by a layer of native niobium oxide ($Nb_2O_5$) - which then can easily form Nb hydride compounds at suitable temperatures and concentrations (*20, 21*). Hydrogen atoms can occupy 12 tetrahedral sites of a bcc Nb lattice, and depending on H concentration, can form a variety of different Nb hydrides in the Nb matrix (*20*). The additional microwave dissipation emerges due to Nb hydrides being poor or non-superconductors (*22*), gaining superconductivity only by the proximity effect within the superconducting niobium matrix (*18*). The possible presence and dissipation due to niobium hydrides, well-studied in bulk niobium SRF cavities, has not been so far considered for the 2D superconducting qubits.

In this study, we apply the same techniques used to reveal the niobium hydrides in bulk SRF cavities, to niobium film regions in superconducting qubit test chips fabricated by Rigetti Computing. We combine room and cryogenic temperature atomic force microscopy (AFM), atomic scale high-resolution and scanning transmission electron microscopy (HR-TEM and STEM) on focused ion beam (FIB) prepared lamellae, and the time-of-flight secondary ion mass spectroscopy (TOF-SIMS) to reveal the existence of the niobium hydrides directly in the Rigetti superconducting qubits.

**Results**

In the inset of **Fig. 1(A),** a schematic of the superconducting qubit chip analyzed in this study is displayed with the Si substrate, Al/$AlO_x$/Al Josephson junction area, and Nb film pads labeled accordingly. The surface morphology of the Nb film is first analyzed using atomic force microscopy (AFM) at room and cryogenic temperatures, as shown in **Fig. 1(A-D)**. At room temperature (RT), the surface of Nb is smooth, displaying root-mean-square (RMS) roughness values < 1 nm. However, we observe two distinct morphological features that arise during cooling down to 2 K and warming up back to 300 K. Irregularly shaped structures of ~500 nm lateral diameter and 10-20 nm of height appear on the surface of the Nb films as the temperature approaches 200 K from the first cooling down to 2 K and persist throughout warming up to 250 K as shown in **Fig. 1(A,B)**. These features completely disappear at 300K. Additional topographical features with the size on the order of ~2 nm are observed to emerge in some of the Nb grains as well from initial cooldown below 50 K, **Fig. 1(C,D)**. In the case of bulk Nb SRF cavities, similar surface features have been previously detected at cryogenic temperatures and have been shown to result from the precipitation of Nb hydrides (*17, 23*).

TOF-SIMS was used to probe the chemical identity of these surface features. Specifically, we detect an appreciable level of hydrogen at the surface from room temperature depth profile measurements directly performed on several Nb contact pad regions from the same qubit device, **Fig. 2(A)**. We find that the H-/Nb- signal, which represents a measure of the free hydrogen concentration present in the parent Nb matrix, is maximum immediately beneath the niobium oxide, and then decays to a much lower level within the first ~10 nm from the surface. The presence of this hydrogen at the surface corroborates with previous findings in bulk Nb cavities and can potentially drive Nb hydride formation in Nb thin film geometries as well. Specifically, compared to a bulk Nb cavity, we observe that the H-/Nb- signal in the top 25 nm closest to the surface is larger in the contact pad (**Fig. 2(B)**). In addition to hydrogen, a significant concentration of other impurities is observed – examples of the observed O-/Nb- and C-/Nb- signals, which are also decaying on a longer length scale than the H-/Nb- signal, are shown in **Fig. 2(A)**.

To investigate these features further, we employed scanning/transmission electron microscopy. An annular dark-field (ADF)-STEM image in **Fig. 3** displays the cross-section of the Nb film on Si and reveals columnar structures of Nb grains. In order to understand the hydride formation process better, Nb thin films from superconducting qubits were analyzed both before and after cooling down to 103 K and warming back to RT. HR-TEM image and their associated fast Fourier transforms (FFT) in **Fig. 4(A, B)** show that the typical lattice structures of bcc Nb on [110] zone axis before cooling down. The schematic of atomic structure of bcc Nb with hydrogen impurities are illustrated in **Fig. 4(C)**, and shows that the hydrogen atoms are randomly distributed in Nb lattice and form a bcc Nb solid solution with hydrogen impurities.

Subsequently, the TEM sample is cooled down to 103 K for 5 hrs and then warmed up to 300 K. After warming up, Nb thin films are analyzed using HR-TEM and we observe that β-NbH precipitates are present within some of the Nb grains. We estimate the concentration of these hydride precipitates to be roughly 0.1-0.5 vol.%. Assuming the hydrogen concentration in top 10 nm of the film closest to the surface is ~2 at.% and other regions are negligible, the average hydrogen concentration in ~170 nm thick Nb film is ~0.1 at.%. If the entirety of these free hydrogen atoms participate in the formation of Nb hydride precipitates, the volume fraction of β-NbH precipitates in Nb is ~0.2 vol.%, which is in qualitative agreement with our observations. The HR-TEM image and corresponding FFT pattern of the Nb grain with hydride precipitates are shown in **Fig. 4(D, E).** The lattice structure of bcc Nb on [110] zone axis with β-NbH precipitates in Nb matrix is presented in the HRTEM image and reflections from β-NbH in the FFT are indicated by yellow arrows. The inset in **Fig. 4(E)** presents the magnified HR-TEM image of β-NbH precipitates, showing the superlattice structure caused by the periodic arrangement of hydrogen in β-NbH within the Nb matrix. The atomic structure of β-NbH is illustrated in the schematic of **Fig. 4(F)**.

The distribution of the β-NbH precipitates in Nb is analyzed and displayed in **Fig. 5**. Specifically, an inverse FFT image provides a map of the β-NbH precipitates in the Nb matrix of **Fig. 5 (A, B)**. This image was constructed by selectively filtering signal associated with the reflections from β-NbH in the FFT image indicated by yellow arrows in **Fig. 4 (D).** We find that the shapes of β-NbH precipitates are irregular and the interface between β-NbH precipitates and bcc Nb is not well-defined. This may imply that β-$NbH_x$ precipitates have compositional variation ($0.5 < x < 1$) and/or defects such as hydrogen vacancies. Magnified images of the boundary between bcc Nb matrix and β-NbH precipitates are seen in **Fig. 5 (C, D)**.

## Discussion

There are several possibilities for the causes of the hydrogen incorporation and subsequent hydride formation in the qubit Nb films. Firstly, hydrogen atoms may be incorporated during the deposition process due to residual hydrogen in the sputter chamber (*31, 33*). Furthermore, as studied in detail for bulk niobium, any chemical (*24*), annealing (*25, 26*), or mechanical polishing (*27*) steps in the manufacturing process performed in a hydrogen-containing environment, which eliminates the protective $Nb_2O_5$ layer on top of Nb, can lead to hydrogen incorporation in the underlying Nb. Further investigation is necessary to pinpoint the particular fabrication steps introducing the hydrogen incorporation into niobium films.

Oxygen, nitrogen and other impurities serve as effective hydrogen trapping centers (*28*), e.g. preventing formation of the large hydrides (*17*) responsible for the "hydrogen Q disease" in low RRR bulk niobium SRF cavities (*29*). Similarly, diffusion of oxygen during the 120°C heat treatments have been found to trap hydrogen and inhibit the nanohydride formation upon cavity cooldown, eliminating the high field Q-slope in high RRR SRF cavities (*30*). Given high concentrations of non-hydrogen impurities revealed by our TOF-SIMS results, the formation of the Nb hydrides in Nb film parts of the qubit chip is somewhat unexpected. One possible explanation is the presence of a much larger concentration of crystalline defects – e.g. dislocations, vacancies, grain boundaries - in the films as compared to bulk niobium in SRF cavities. This can both make the hydrogen distribution highly non-uniform, enabling local higher concentration areas, and provide plenty of hydride nucleation sites to facilitate precipitation.

It is important to note that the hydride morphology may change with subsequent cooling cycles, as studied in detail in bulk Nb (*17*). This can provide a potential explanation for observations relating to changes in the qubit coherence times following each cooldown, as well as the effect of "aging" from the accumulated number of cooldowns. In the case of superconducting RF Nb cavities, for instance, incorporation of hydrogen within interstitial sites leads to the so-called "Q slope" where the NbH proximity coupled to the surrounding Nb matrix leads to a strong degradation of the quality factor at high fields (*16, 18*). Additionally, because this proximity exhibits a smaller superconducting gap than the surrounding Nb, it may also serve as a sink for quasiparticles and in turn, diminish energy relaxation coherence times near the operating temperature. Additionally, the presence of these precipitates throughout the entirety of the film suggests that simply increasing or decreasing film thickness is insufficient for reducing the impact of this decoherence source.

While the volume fraction of hydride in Nb thin film superconducting qubits of the current study is limited (0.1-0.5 vol.%), we note that the volume fraction of hydride precipitates could increase substantially if the hydrogen impurity in the Nb thin film is not well controlled. For comparison, cryo-TEM analysis is performed on mechanically polished bulk Nb sample for RF cavities with ~2 order of magnitude higher hydrogen concentration based on SIMS analysis, **Fig. 2(B).** In this Nb sample, we observe more significant hydride precipitation during cooling down along the whole Nb grains, which is roughly estimated to be 3-6 vol.% of $\zeta$-$NbH_{0.5}$ assuming the average hydrogen concentration in the bulk Nb sample is 1-2 at.%, **Fig. S1**. This indicates that the concentration of free hydrogen in Nb plays a key role in the hydride precipitation in Nb and controlling the hydrogen impurities in Nb during film growth and fabrication is important to suppress the hydride formation in Nb thin film superconducting qubits.

We have also investigated the possible effect of Ga ion beam on the formation of hydrides in Nb during the sample preparation process using bulk Nb samples. There have been reports that Ga ion beam introduces hydrogen and large distinct shaped hydride phases could form in Ti (*32*) and Zr (*33*) during the TEM sample preparation using FIB. However, Ti and Zr have a higher affinity to hydrogen than Nb and we could not observe hydride formation in the TEM foils of bulk Nb after TEM sample preparation using FIB with the same parameters. Therefore, we conclude that Ga ion beam effect is not significant in Nb foils and Nb hydride phases we observed in the current study are formed during or after the film growth or subsequent chip patterning and fabrication process.

There could be several strategies to mitigate the formation of Nb hydride precipitates in Nb films. Post-annealing of Nb films at above around 600°C, similar to bulk niobium standard processing steps, may degas hydrogen atoms from Nb and reduce the formation of Nb hydride domains (*34, 35*). One potential challenge is that at these higher temperatures Nb silicide can be formed at the Nb/Si interfaces, thus using a sapphire substrate instead may enable a more suitable temperature range of heat treatments. The structure of GBs may also play a role in the formation of Nb hydrides near GBs of Nb (*36, 37*). Further optimization of the microstructure of Nb films such as grain size and impurities may provide an additional pathway to mitigate the hydride formation and suppress the hydride-induced decoherence in Nb superconducting planar resonators and superconducting qubits. Such an optimization may be performed by exploring various state-of-the-art niobium film deposition techniques, including those leading to much higher RRR values.

In summary, our findings suggest that the existence of the Nb hydride precipitates in Nb thin films could be a widespread and so far unaccounted for source of the decoherence in superconducting qubits and planar Nb resonators. Niobium hydrides may contribute to both TLS and quasiparticle dissipation, as well as serve as a source of the cooldown-to-cooldown variability and "aging".

**Acknowledgments:** We thank Rigetti Computing and, in particular, Cameron Kopas and the Rigetti chip design and fabrication teams for the development and manufacturing of the qubit devices used in the reported experimental study. We also thank Charlene Wilke and Dr. Xiaobing Hu for the valuable support and advice for the cryo-TEM experiments.

We also acknowledge helpful discussions with Carlos Gerardo Torres Castanedo, Drs. Paul Chandrica Masih Das, Dominic Pascal Goronzy, Profs. Mark Hersam and David N Seidman.

**Funding:** This material is based upon work supported by the U.S. Department of Energy, Office of Science, National Quantum Information Science Research Centers, Superconducting Quantum Materials and Systems Center (SQMS) under contract number DE-AC02-07CH11359. This work made use of the EPIC, Keck-II, and/or SPID facilities of Northwestern University's NU*ANCE* Center, which received support from the Soft and Hybrid Nanotechnology Experimental (SHyNE) Resource (NSF ECCS-1542205); the MRSEC program (NSF DMR-1121262) at the Materials Research Center; the International Institute for Nanotechnology (IIN); the Keck Foundation; and the State of Illinois, through the IIN.

**Author contributions:** A. R. and A. G. designed and guided the experiments and data analysis. M. J. R. provided superconducting qubit chips. Z. S. performed cryo-AFM investigations, J. L. performed FIB/TEM investigations, and A. A. M. performed TOF-SIMS investigations. J. L., Z. S., A. A. M., and A.R. wrote the manuscript with input from all authors.

**Competing interests:** The authors declare that they have no competing interests.

**Data and materials availability:** All data are available in the main text or the supplementary materials.

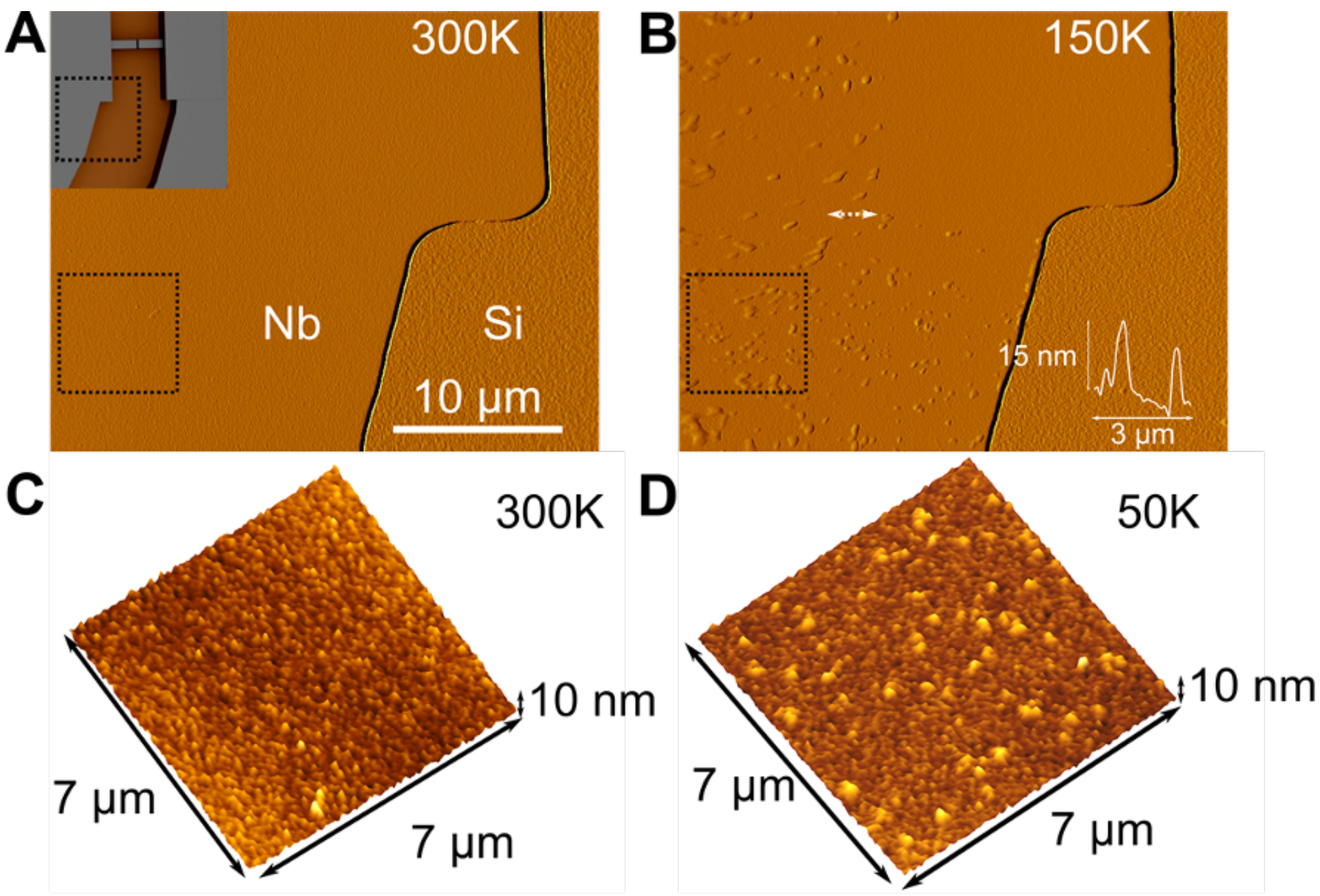

**Fig. 1** (**A**) Inset - schematic of the Nb qubit pads area, cryo-AFM area of analysis is shown by dashed rectangle; AFM surface images of Nb films and Nb/Si boundary at room temperature, and (**B**) at T=200 K (from initial cooling down to 2 K), with the formation of the unknown structures on the surface of Nb clearly observed; the line profile through the structures at the white arrow is shown in the inset on the right, indicating the height of the order of 15 nm. The higher magnification surface topology of Nb films from the dashed rectangles in (A) and (B) is shown in respectively (**C**) RT and (**D**) 50 K (during initial cooldown to 2 K) - with further smaller scale formations apparent.

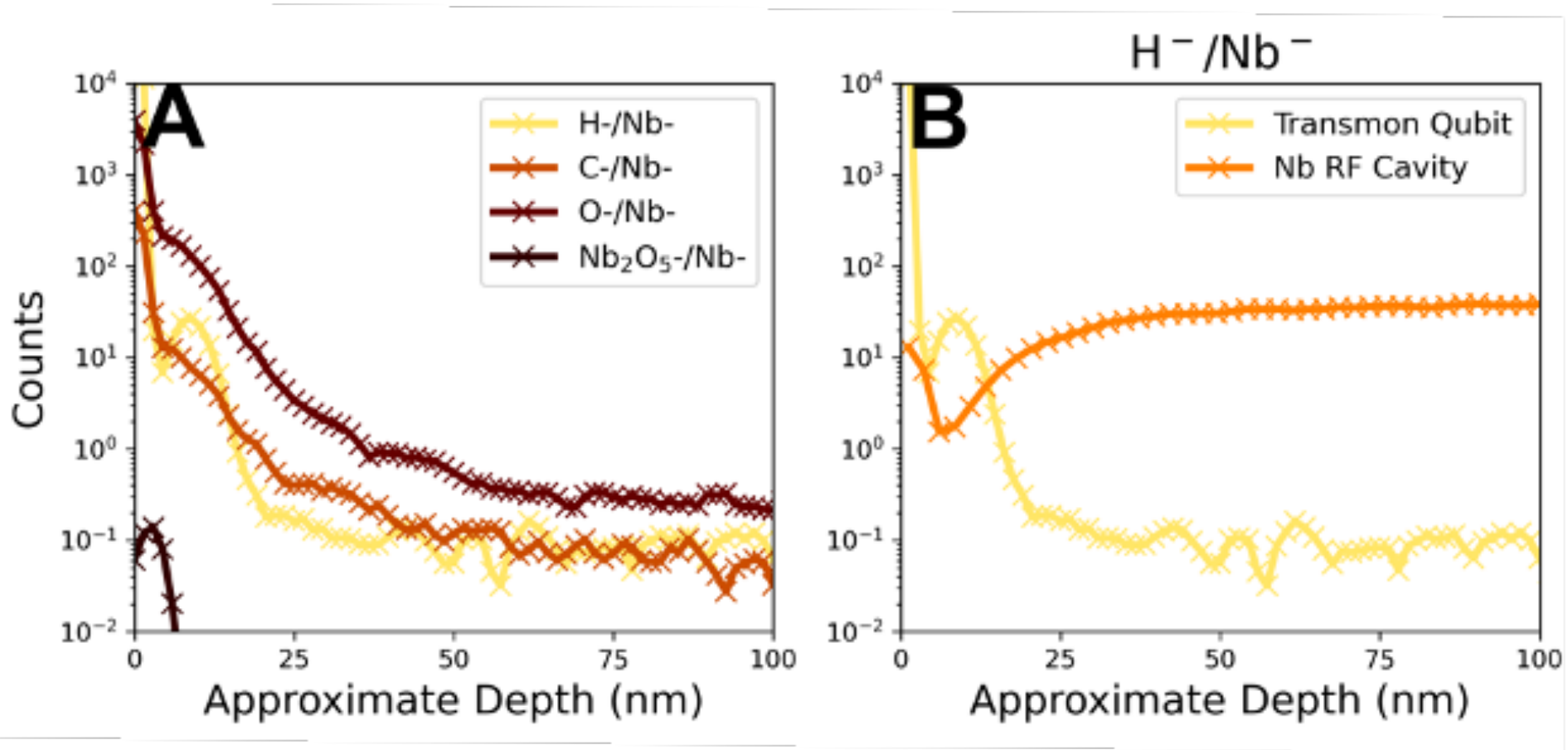

**Fig. 2** (A) SIMS depth profiles of O, H, C impurities and the $Nb_2O_5$ surface oxide layer in Nb film in the superconducting qubit after the fabrication process. (B) The comparison of hydrogen impurity in transmon qubit and bulk Nb.

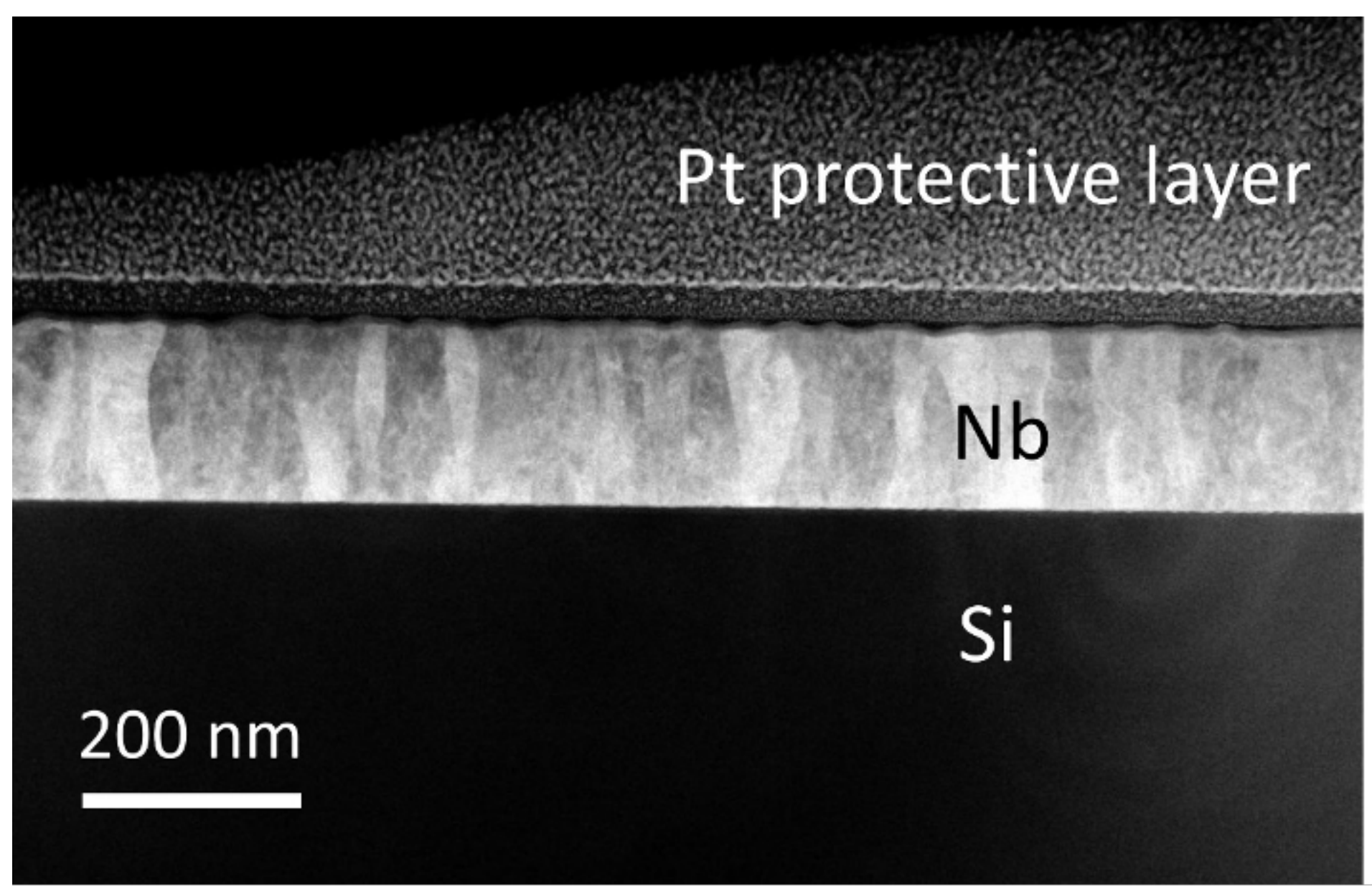

**Fig. 3** Annular dark-field (ADF)-STEM image of Nb/Si cross-section of Nb resonator on Si.

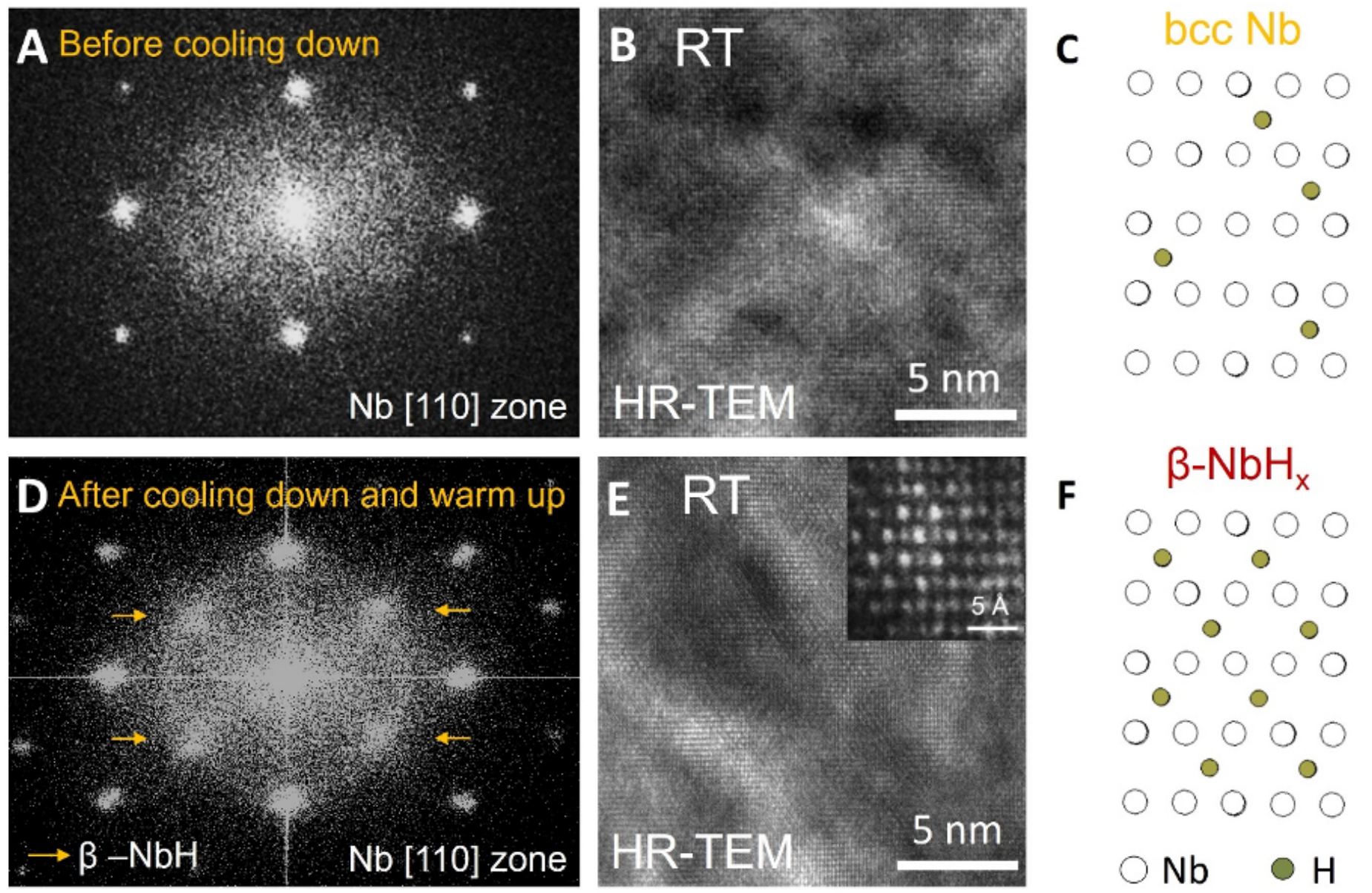

**Fig. 4** HR-TEM image and corresponding FFT images of Nb thin films before and after cooling down to 103 K and warming back to room temperature (298 K) are displayed. (A-B) Before cooling down, HR-TEM and corresponding FFT images show the lattice structure of bcc Nb on [110] zone axis. (C) Atomic structures of bcc Nb and β-NbH are illustrated on bcc Nb [110] zone. In bcc Nb, hydrogen atoms are randomly distributed in the lattice. (D-E) After cooling down to 103 K and warming up, superlattice reflections from β-NbH appear, which are denoted by yellow arrows in the FFT image, Fig. 4(D). Inset in Fig.4(E) reveals the superlattice structure of β-NbH in HR-TEM image caused by the periodic array of hydrogen in Nb lattice. The HR-TEM image is taken around the center of the Nb film between the surface and Nb/Si interface. (F) The schematic of the atomic structure of β-NbH illustrates that when hydrogen atoms are sitting in the Nb lattice in order, β-NbH precipitates are formed.

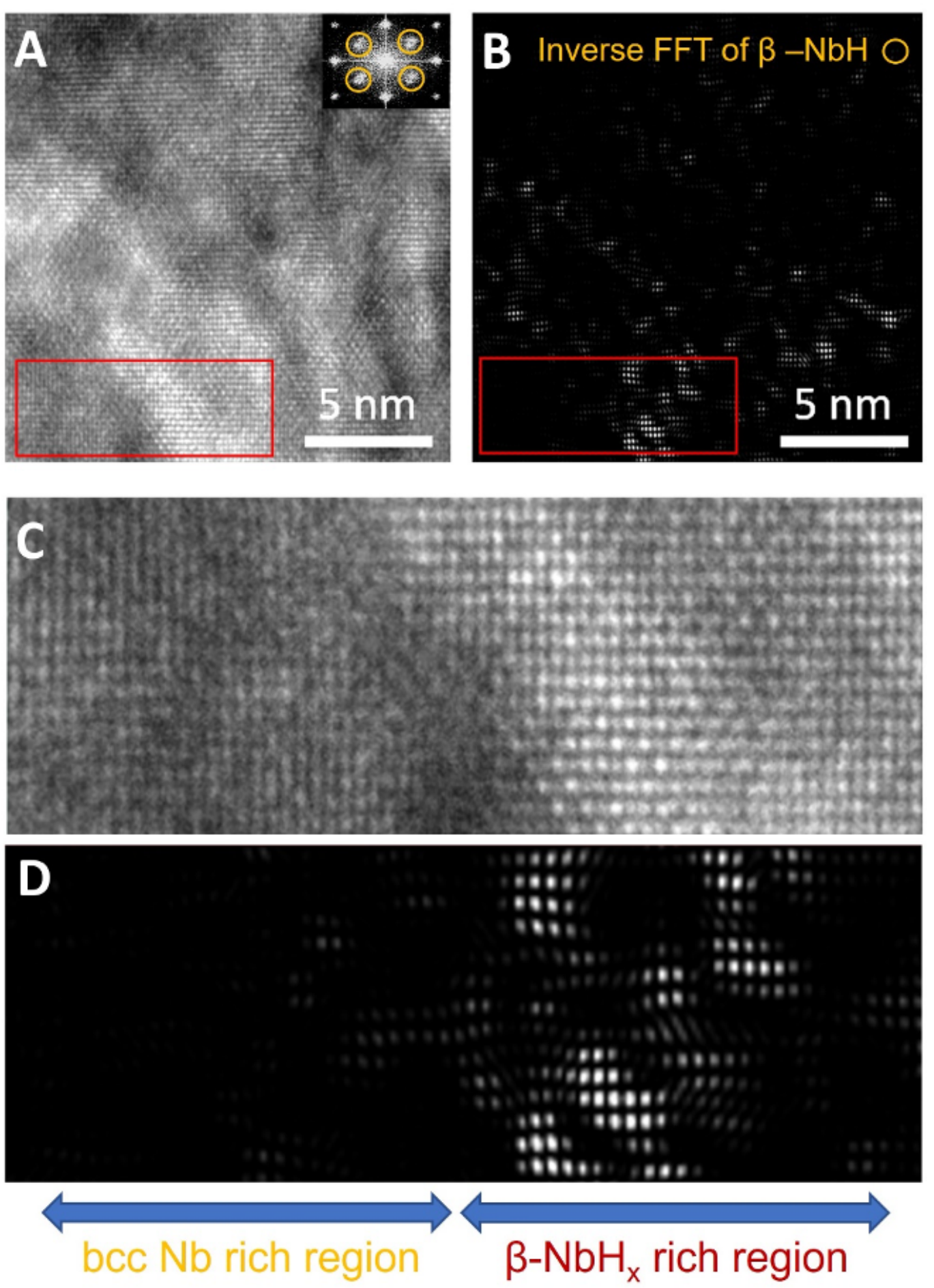

**Fig. 5** Distributions of β-NbH in bcc Nb are illustrated in HR-TEM and inverse FFT images. (A) HR-TEM image displays the lattice structure of β-NbH embedded in a bcc Nb grain. FFT image of the corresponding HR-TEM image reveals the superlattice reflection caused by β-NbH, denoted by yellow circles. (B) Inverse FFT from the β-NbH reflections illustrate the distribution of β-NbH in the Nb grain in nanoscale. (C) The magnified HR-TEM and (D) inverse FFT images from red squares in (A, B) illustrate the boundary between bcc Nb rich region and β-NbH rich region.

Supplementary Materials

Materials and Methods

Superconducting qubit test devices with Nb resonators and $Al/AlO_x/Al$ Josephson junctions have been fabricated on a Si (100) substrate at Rigetti Computing based on the published nanofabrication procedure (*38*). The chip containing five superconducting qubits, readout resonators, and connecting lines has been used for the studies.

To search for niobium hydride formation, several complementary approaches were used. Firstly, to study the internal niobium structure throughout the films, focused ion beam has been used to prepare several TEM lamellae for subsequent room and cryogenic temperatures studies. Electron diffraction has been used to search for and confirm the presence of niobium hydride phases in the niobium matrix.

Secondly, the atomic force microscopy (AFM) using the attoCube system, incorporated with Quantum Design 9T (tesla) PPMS (physical property measurement system) has been performed to directly look at various areas of the superconducting qubit chip at the range of temperature from room to 2 K, in helium gas environment. AFM scanning in standard contact mode with a nanoseonsors PPP (pointprobe plus) tip allows a plan view of about 18×18 μm to be observed throughout the changing temperatures to search for morphological changes on the surface caused by the niobium hydride formation, similar to refs. (*16, 19, 39*).

Scanning electron microscopy (SEM) imaging of the superconducting qubit and TEM sample preparation were carried out with Helios Nanolab 650. TEM lamellae were prepared using 30 kV Ga ion beam and finely polished by 5 kV and 2 kV Ga ion beams in order to obtain high electron transparent imaging as well as to remove the damaged surface on the Nb foils. Annular dark-field scanning transmission electron microscopy (ADF-STEM) and HR-TEM were performed on JEOL ARM200CF electron microscope operated at 200 kV. The microscope was equipped with a Cold FEG source and probe aberration corrector. ADF images were acquired using a convergence semi-angle of 21 mrad and collection angles of 68-260 mrad. Gatan $LN_2$ single-tilt TEM cold stage was used to cool the foils down to cryogenic temperature (106 K), which is confirmed to be sufficient for β, ε, ζ, λ–NbH phase (*19*).

To provide further insight into the origin of the niobium hydride emergence, we have used a dual beam time-of-flight secondary ion mass spectrometry (TOF-SIMS) – IONTOF 5 - to analyze the concentration and depth distribution of hydrogen in Nb parts of the chip. Secondary ion measurements were performed using a liquid bismuth ion beam (Bi+) and a cesium ion gun with an energy of 500 eV was used for sputtering the surface for depth profile measurements.

Supplementary text

For comparison, we also analyzed hydride precipitation in bulk Nb sample of the same kind used for superconducting radiofrequency (SRF) cavities. Heavy hydrogen loading of the sample was performed using the conventional mechanical surface polishing, followed by water-solution based colloidal silica VibroMet polishing. About two orders of magnitude higher hydrogen concentration is achieved this way, as compared to the concentrations within Nb thin films in the superconducting qubits we studied. TEM sample has been prepared from the surface of such a mechanically polished Nb by FIB lift-out and BF-TEM image in **Fig. S1(A)** display bi-crystal Nb grain structure. We observe the β-NbH precipitation already present in the mechanically polished bulk Nb at room temperature, as seen in the electron diffraction pattern of **Fig. S1(B, C)**. After cooling down to 103 K, the electron diffraction pattern indicates that another type of hydride precipitates with lower hydrogen concentration - ζ-$NbH_{0.5}$. It is formed along the Nb foils, **Fig. S1(D)**. The TEM sample is warmed up and left in the air at RT for 6 months. Then, we analyze the Nb grain with HR-TEM and still observe β-NbH precipitate at RT, but ζ-$NbH_{0.5}$ precipitates disappeared, **Fig. S1(E)**. In the bulk Nb with ~2 order of magnitude higher hydrogen concentration, we observe much higher volume fraction of hydride precipitations and their systematic discrepancies on precipitation/dissolution during cooling down and warming up cycle.

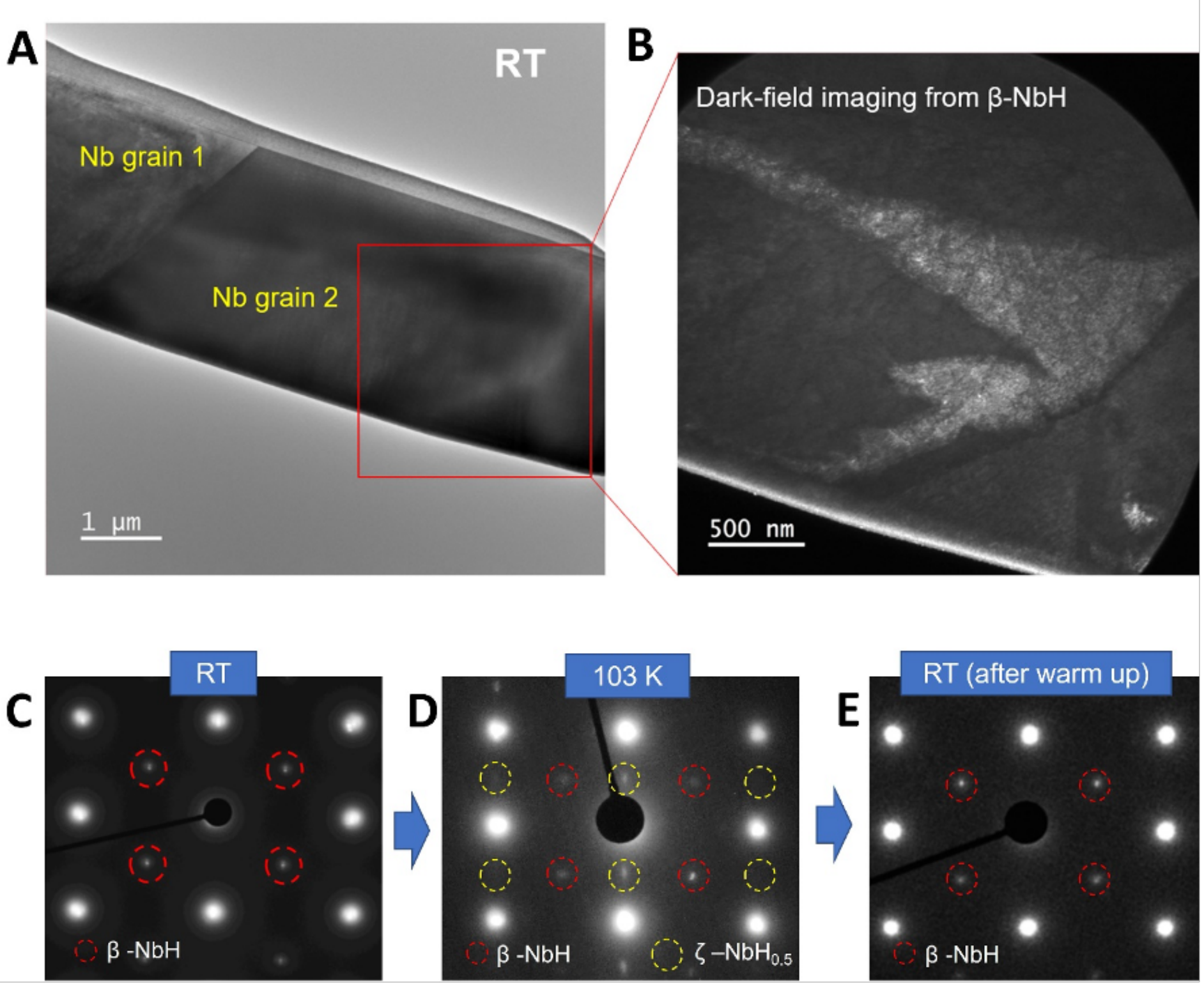

**Fig. S1** (A) BF-TEM image of hydrogen-loaded bulk Nb coupon is shown. It contains ~2 order of magnitude higher hydrogen concentration than Nb thin film in superconducting qubits. (B) Dark-field image from the reflections of the β-NbH phase show the irregular shape β-NbH in Nb at RT. (C-E) Diffraction patterns of the Nb grain on [110] zone axis are collected at RT, at 103 K and after warming up back to 300 K, systematically. It illustrates the precipitation and dissolution of ζ-$NbH_{0.5}$ during cooling down and after warming up.

## References and Notes

1. C. Müller, J. H. Cole, J. Lisenfeld, Towards understanding two-level-systems in amorphous solids: insights from quantum circuits. *Rep. Prog. Phys.* **82**, 124501 (2019).
2. J. Zmuidzinas, Superconducting microresonators: Physics and applications. *Annu. Rev. Condens. Matter Phys.* **3**, 169-214 (2012).
3. J. M. Martinis *et al.*, Decoherence in Josephson qubits from dielectric loss. *Phys. Rev. Lett.* **95**, 210503 (2005).
4. D. P. Pappas, M. R. Vissers, D. S. Wisbey, J. S. Kline, J. Gao, Two level system loss in superconducting microwave resonators. *IEEE Transactions on Applied Superconductivity* **21**, 871-874 (2011).
5. A. A. Murthy *et al.*, Potential Nanoscale Sources of Decoherence in Niobium based Transmon Qubit Architectures. *arXiv preprint arXiv:.08710*, (2022).
6. A. Premkumar *et al.*, Microscopic Relaxation Channels in Materials for Superconducting Qubits. *arXiv:.02908*, (2020).
7. M. Altoé *et al.*, Localization and reduction of superconducting quantum coherent circuit losses. *arXiv:.07604*, (2020).
8. A. Romanenko, D. Schuster, Understanding quality factor degradation in superconducting niobium cavities at low microwave field amplitudes. *Phys. Rev. Lett.* **119**, 264801 (2017).
9. A. Megrant *et al.*, Planar superconducting resonators with internal quality factors above one million. *Appl. Phys. Lett.* **100**, 113510 (2012).
10. C. T. Earnest *et al.*, Substrate surface engineering for high-quality silicon/aluminum superconducting resonators. *Superconductor Science Technology* **31**, 125013 (2018).
11. J. Wenner *et al.*, Surface loss simulations of superconducting coplanar waveguide resonators. *Applied Physics Letters* **99**, 113513 (2011).
12. A. Romanenko *et al.*, Three-Dimensional Superconducting Resonators at $T< 20$ mK with Photon Lifetimes up to $\tau= 2$ s. *Physical Review Applied* **13**, 034032 (2020).
13. K. Serniak *et al.*, Hot nonequilibrium quasiparticles in transmon qubits. *Phys. Rev. Lett.* **121**, 157701 (2018).
14. A. P. Vepsäläinen *et al.*, Impact of ionizing radiation on superconducting qubit coherence. **584**, 551-556 (2020).
15. C. D. Wilen *et al.*, Correlated charge noise and relaxation errors in superconducting qubits. *Nature* **594**, 369-373 (2021).
16. F. Barkov, A. Romanenko, A. Grassellino, Beams, Direct observation of hydrides formation in cavity-grade niobium. *Physical Review Special Topics-Accelerators* **15**, 122001 (2012).
17. F. Barkov, A. Romanenko, Y. Trenikhina, A. Grassellino, Precipitation of hydrides in high purity niobium after different treatments. *J. Appl. Phys.* **114**, 164904 (2013).

18. A. Romanenko, F. Barkov, L. Cooley, A. Grassellino, Proximity breakdown of hydrides in superconducting niobium cavities. *Supercond. Sci. Technol.* **26**, 035003 (2013).

19. Y. Trenikhina, A. Romanenko, J. Kwon, J.-M. Zuo, J. Zasadzinski, Nanostructural features degrading the performance of superconducting radio frequency niobium cavities revealed by transmission electron microscopy and electron energy loss spectroscopy. *J. Appl. Phys.* **117**, 154507 (2015).

20. B. Makenas, H. Birnbaum, in *Perspectives in Hydrogen in Metals*. (Elsevier, 1986), pp. 523-535.

21. T. Schober, H. Wenzl, in *Hydrogen in Metals II*. (Springer, 1978), pp. 11-71.

22. G. Rauch, R. Rose, J. Wulff, Observations on microstructure and superconductivity in the Nb-H system. *Journal of the Less Common Metals* **8**, 99-113 (1965).

23. Z. Sung, A. Romanenko, M. Martinello, A. Grassellino, in *Proceedings of the International Conference on RF Superconductivity SRF*. (2019).

24. T. Higuchi, K. Saito, Y. Yamazaki, T. Ikeda, S. Ohgushi, in *Proceedings of the 10th Workshop on RF Superconductivity*. (2001), pp. 427-430.

25. K. Faber, H. Schultz, Hydrogen contamination in tantalum and niobium following UHV-degassing. *Scr. Metall. Mater.* **6**, 1065-1070 (1972).

26. S. Isagawa, Hydrogen absorption and its effect on low-temperature electric properties of niobium. *J. Appl. Phys.* **51**, 4460-4470 (1980).

27. T. Higuchi, K. Saito, *Proceedings of the 11th Workshop on RF Superconductivity*, 572–578 (2003).

28. D. C. Ford, L. D. Cooley, D. N. Seidman, Suppression of hydride precipitates in niobium superconducting radio-frequency cavities. *Supercond. Sci. Technol.* **26**, 105003 (2013).

29. K. Saito, H. Miwao, T. Suzukie, Behavior of Electropolished Niobium Cavities Under Different Cooldown Conditions. *Proceedings of the 5th Workshop on RF Superconductivity*, 665-679 (1991).

30. A. Romanenko, Y. Trenikhina, M. Martinello, D. Bafia, A. Grassellino, First direct imaging and profiling TOF-SIMS studies on cutouts FOM cavities prepared by state-of-art treatments. *Proceedings of the 19th International Conference on RF Superconductivity*, THP014 (2019).

31. A. A. Murthy *et al.*, TOF-SIMS analysis of decoherence sources in superconducting qubits. *Appl. Phys. Lett.* **120**, 044002 (2022).

32. Y. Chang *et al.*, Ti and its alloys as examples of cryogenic focused ion beam milling of environmentally-sensitive materials. *Nature communications* **10**, 1-10 (2019).

33. H. Shen, X. Zu, B. Chen, C. Huang, K. Sun, Direct observation of hydrogenation and dehydrogenation of a zirconium alloy. *J. Alloys Compd.* **659**, 23-30 (2016).

34. D. Bafia *et al.*, in *19th International Conference on RF Superconductivity (SRF'19), Dresden, Germany, 30 June-05 July 2019*. (JACOW Publishing, Geneva, Switzerland, 2019), pp. 586-591.

35. A. Grassellino *et al.*, Unprecedented quality factors at accelerating gradients up to 45 MVm− 1 in niobium superconducting resonators via low temperature nitrogen infusion. *Supercond. Sci. Technol.* **30**, 094004 (2017).

36. P. Garg *et al.*, Revealing the role of nitrogen on hydride nucleation and stability in pure niobium using first-principles calculations. *Supercond. Sci. Technol.* **31**, 115007 (2018).

37. Z.-H. Sung *et al.*, Development of low angle grain boundaries in lightly deformed superconducting niobium and their influence on hydride distribution and flux perturbation. *J. Appl. Phys.* **121**, 193903 (2017).

38. A. Nersisyan *et al.*, in *2019 IEEE International Electron Devices Meeting (IEDM)*. (IEEE, 2019), pp. 31.31. 31-31.31. 34.

39. Z. Sung, A. Romanenko, M. Martinello, A. Grassellino, in *International Conference on RF superconductivity*. (2019).